
\input harvmac
\noblackbox
\tolerance=1600
\Title{\vbox{\baselineskip12pt\hbox{HUTP-91/A048}}}{
Large-Small Equivalence in String Theory }

\centerline{Eva Silverstein}
\bigskip\centerline{Lyman Laboratory of Physics}
\centerline{Harvard University}\centerline{Cambridge, MA 02138, USA}

\vskip .3in
The simplest toroidally compactified string theories
exhibit a duality between small and large radii: compactification
on a circle, for example, is invariant under
$R\rightarrow 1/(2R)$. Compactification on more
general Lorentzian lattices (e.g. toroidal compactification
in the presence of background metric, antisymmetric tensor, and
gauge fields) yields theories for which a large-small equivalence
is not so simple.
Here an equivalence is
demonstrated between large and small
geometries for all toroidal compactifications.
By repeatedly transforming the momentum mode corresponding
to the smallest winding dimension to another mode on the
lattice, it is possible to increase the volume to
exceed a finite lower bound.
\Date{11/91} 

\newsec{Introduction}
It has been known for some time that the simplest toroidal string
compactification exhibits an equivalence between small and large
background geometries.  That is, as pointed out in
\ref\Arpop{K. Kikkawa and M. Yamasaki, Phys Lett. 149B (1984) 357\semi
          N. Sakai and I. Senda, Prog. Theor. Phys. 75 (1986) 692} and
later expanded upon in \ref\Arpop{V.P. Nair, A. Shapere,
A. Strominger and F. Wilczek, Nucl. Phys. B287 (1987) 402\semi
P. Ginsparg and C. Vafa, Nucl. Phys. B289 (1987) 414\semi
B. Sathiapalan, Phys. Rev. Lett. 58 (1987) 1597\semi
R. Dikgraaf, E. Verlinde and H. Verlinde, CMP 115, 649-90 (1988)\semi
A.A. Tseytlin, "Duality and Dilaton" preprint JHU-TIPAC-91008\semi
R. Brandenberger and C. Vafa, Nucl. Phys. B316 (1988) 391\semi
A.A. Tseytlin and C. Vafa, "Elements of String Cosmology",
preprint HUTP/AO49}
and [4], the
presence of winding states leads to an $R\rightarrow 1/(2R)$ invariance of a
theory compactified on a circle without any background fields; this
generalizes easily to a G to $1/4G^{-1}$ invariance for a theory
compactified on a torus given by the metric G.
($G_{ij}=e_i\cdot e_j$ where the $e_i$ form a basis for the winding
vectors.)

This raises the question of how generally this phenomenon
occurs in arbitrary toroidal compactifications of
string theory.  To answer this question we must consider
the moduli space of Lorentzian lattice
compactifications spanned by the background
metric, antisymmetric tensor, and gauge field (Wilson lines).
The worldsheet action is
$$S=\int{G_{ij}\partial_{\alpha}X^{i}\partial^{\alpha}X^{j}
+ \epsilon^{\alpha\beta}B_{ij}
\partial_{\alpha}X^{i}\partial_{\beta}X^{j}
+ \epsilon^{\alpha\beta}A_i^J
\partial_{\alpha}X^i\partial_{\beta}X^J}$$
As described in \ref\clos{K.S. Narain, M.H. Sarmadi and E. Witten,
Nucl.Phys. B279 (1986) 369}, the left- and right-moving momenta
are
$$p_L=\bigl(P+A^Tn, G^{-1}\left(m/2-Bn-(1/4)AA^Tn-(1/2)AP\right)+n\bigr)$$
and
$$p_R=\bigl(G^{-1}\left(m/2-Bn-(1/4)AA^Tn-(1/2)AP\right)-n\bigr)$$

where n,m $\epsilon Z^d$ are the winding and momentum vectors
and P is a vector in the 16-dimensional root lattice of the
gauge group.
With general B and A values,
interchanging $G$ and $1/4G^{-1}$ does not always yield a new state on
the same lattice or even preserve the spectrum of
$p_L^2+p_R^2$ values, as would be necessary for an invariance of
this simple form.  As discussed in
\ref\clos{A. Shapere and F. Wilczek, Nucl. Phys. B320 (1989) 669}, the moduli
space
of Lorentzian lattice compactifications contains discrete
equivalences given by O(16+d,d,Z) transformations on ($p_L$,$p_R$), which
leave the lattice invariant but which can be obtained
equivalently by changing the background fields.
(Here d is the number of compactified spatial dimensions.) That is,
the moduli space of compactifications is
O(16+d,d)$\big/$O(16+d)xO(d)xO(16+d,d,Z).  One
example is the transformation taking $B_{ij}$ to
$B_{ij}+1$; another is the well-known so-called duality
transformation taking 2(B + G) to its inverse (discussed in
[4] and \ref\clos{A. Giveon, E. Rabinovici, and G. Veneziano,
Nucl. Phys. B322 (1989) 167}).

Thus we must
investigate whether there exists some O(16+d,d,Z) transformation
taking small G to large G.  First we must specify precisely
what we mean by the size of the compactified space.  One condition
for small-large duality
requires the volume det G to have some finite effective lower
bound.  A stronger condition is that
in addition the winding lengths $n^TGn$ should themselves
all exceed some lower bound.  As discussed in section 3, it can
be shown that the latter implies the former: in any dimension,
if all winding lengths exceed some minimum, the volume must
also have a lower bound.

In this paper we show that
the compactified volume $\sqrt{\rm{det} G}$ can be raised to exceed
a finite volume; we can make the further statement that
if the lengths of the windings cannot all be made to exceed
a finite lower bound, then we can transform the volume to $\infty$.
In two dimensions with A=0 we will see explicitly that the winding
lengths can be raised to exceed a lower bound.  These results suggest
that winding lengths can be raised in general; further work is
necessary to show this for all cases.
We will begin by discussing the d=2 case without Wilson lines in
a way that we will then generalize to higher dimensions.
Finally, we will demonstrate the result with Wilson lines included.
\newsec{Small-Large Duality for $d = 2$ and $A=0$}

As discussed in [4], the generalization of the interchange
of momentum and winding vectors ($n\leftrightarrow m$)
is $2(G+B)\rightarrow 1/2(G+B)^{-1}$.  This transformation
fails to transform all small metrics to large, as the
following example demonstrates.  When $G=RI$, we have
$$2\left(\matrix{R^2&-B\cr
               B&R^2\cr}\right)\rightarrow
{1\over{2(R^4+B^2)}}\left(\matrix{R^2&B\cr
                                  -B&R^2\cr}\right)$$
under this transformation (here $B=-B_{12}$).  Thus as
$R^2\rightarrow 0$ for $B\ne 0$ it is transformed
to another nearly vanishing radius.

It is possible, taking into account other available
O(2,2,Z) transformations, to satisfy both conditions
for small-large duality mentioned in section 1.
As described in [3], we can raise the volume det G for $d=2$ using
$SL(2,Z){\rm x} SL(2,Z) {\it=} SO(2,2,Z)$ transformations on the two complex
parameters $$\rho = 2\left(B + i\sqrt{\rm det G}\right)$$ and
$$\tau = {G_{12}\over{G_{11}}} + {i\sqrt{\rm det G}\over{G_{11}}}$$
We simply need to transform $\rho$ to its fundamental domain
for which $2{\rm det}G \ge \sqrt{3}/2$ (see fig. 1).
It is not too hard to see, however, that $SL(2,Z){\rm x} SL(2,Z)$
transformations do not suffice to raise the
winding lengths of the two-dimensional torus:
transforming $\tau$ by an SL(2,Z) transformation merely
reparameterizes the torus (i.e. changes its basis);
transforming $\rho$ does not affect
the shape of the torus, just its volume.
Consequently, since the fundamental
domain contains only one point in the orbit of each $\rho$,
a torus with a small winding
situated in $\rho$'s fundamental domain cannot be transformed
to one with larger windings by SL(2,Z)xSL(2,Z).  To illustrate this,
consider a diagonal torus with $R_1=\epsilon << 1$ and $R_2=1/\epsilon$.
For this torus, det G = 1, putting it in $\rho$'s fundamental
domain.  Here Im $\tau = R_1/R_2$; transforming $\rho$
leaves this ratio fixed.  Since $\rho$ is already in its
fundamental domain, the volume cannot be further increased.
Thus $R_1$ will stay small as long as we confine
ourselves to SL(2,Z)xSL(2,Z).  If, however, we interchange $\rho$
and $\tau$ (this is equivalent to exchanging the first component
the momentum and winding modes, as will be seen in section 3),
we find
$$G\rightarrow\left(\matrix{1/(4\epsilon^2)&B/(2\epsilon^2)\cr
                            B/(2\epsilon^2)&(1+B^2)/\epsilon^2\cr}
\right)$$
and
$$n^TGn\rightarrow (n_1 + (2B)n_2)^2/(4\epsilon^2)
 + n^2_2/\epsilon^2 \ge 1/(4\epsilon^2)$$

Thus we find that to raise winding lengths in $d=2$ we must
make use of the rest of O(2,2,Z), namely the $Z_2$ interchange of
$\rho$ and $\tau$.  First we show that this transformation
provides a different way to raise the volume from the
SL(2,Z)xSL(2,Z) transformations.
Given a $d=2$ torus G, reparameterize G
if necessary to form a basis out of its two smallest winding
lengths.  The SL(2,Z) generators $\tau\rightarrow{\tau + 1}$,
which takes $e_1 \rightarrow {e_1+e_2}$, and
$\tau\rightarrow{-1/{\tau}}$, which interchanges the 1 and 2
components, provide the necessary transformations for this
reparameterization.  Suppose the smallest squared winding,
$G_{11}$, is less than 1/2.
Then the interchange of $\rho$ and $\tau$, taking det G to
det G/$4G_{11}^2$, increases the volume.  We may repeat this
procedure as long as $G_{11}<1/2$.  Then we have either
${\rm det} G\rightarrow\infty$ or at least $G_{11},G_{22}\ge 1/2$
(i.e. all squared winding lengths $\ge 1/2$,
which is sufficient to put a lower bound on the volume, as
discussed in section 3).  This procedure will generalize to higher d.

In fact, this procedure in $d=2$ gives us a two-step method for
putting $\rho$ in its fundamental domain.  In the basis consisting
of the two smallest windings $e_1\le{e_2}$, the angle between $e_1$
and $e_2$ ($\theta_{12}$) must be between $\pi /3$ and $2\pi /3$;
otherwise $\vert e_2-e_1 \vert < \vert e_2 \vert$, contradicting
the fact that $e_1$ and $e_2$ are the two smallest windings.
Thus once $\tau$ has been transformed to this basis, it falls
in the shaded region in figure 2.
Then $\tau$ can be placed in its fundamental region by
$\tau \rightarrow \tau \pm 1$.  Finally, $\rho \leftrightarrow \tau$
puts $\rho$ in its fundamental region, ensuring that
$2\sqrt{{\rm det}G} > {\sqrt 3 /2}$.

Next we show that a $d=2$ torus can be transformed via
$\rho \leftrightarrow \tau$
into one with all windings bounded below.  To see this we
consider all possible windings $n^TG^{\prime}n$ of the new
metric.  We find
$$G^\prime=\left(\matrix{1/(4G_{11})&B/(2G_{11})\cr
                          B/(2G_{11})&({\rm det}G+B^2)/G_{11}
                             \cr}\right)$$
This gives
$$n^TG^{\prime}n={n_2^2{\rm detG}\over{G_{11}}}
+{(n_1+2n_2B)^2\over{4G_{11}}}$$
We see here that with ${\rm det} G > 3/16$ (i.e. with $\rho$
in its fundamental domain so that
$2\sqrt{{\rm det} G}>\sqrt{3}/2$), interchanging $\tau$
and $\rho$ yields new windings whose squared lengths all exceed
$3/8$: If $n_2 \ne 0$, the squared winding
$e^2 \ge 3/(16G_{11})\ge 3/8$ (for the smallest squared winding
$G_{11}\le 1/2$).
If $n_2 = 0$, we have $e^2 > 1/(4G_{11}) \ge 1/2$.
Thus it takes at most two interchanges of $\rho$ and $\tau$
to transform any torus to one with all windings bounded below.
Unfortunately this argument for putting a lower bound
on the winding lengths does not generalize to all higher d,
although the above argument for the
weaker form of small-large duality does survive in higher
dimensions.

\newsec{Volumes for General d and A=0}

As discussed in [3], $$(1/2)(p_L^2+p_R^2)=
\left(\matrix{n^T&m^T\cr}\right)
\left(\matrix{G-BG^{-1}B&1/2BG^{-1}\cr
              -1/2G^{-1}B&1/4G^{-1}\cr}\right)
\left(\matrix{n\cr
m\cr}\right)$$
Thus to obtain the effect of an O(d,d,Z) transformation on
the background fields, we simply transform the above matrix
accordingly.  That is,
$$\left(\matrix{n \cr
                m \cr}\right) \rightarrow
S \left(\matrix{n \cr
                m \cr}\right)$$
where $S\epsilon O(d,d,Z)$ is equivalent to
$$M(G,B) \rightarrow S^TM(G,B)S$$
where M(G,B) is the matrix of background fields
in the above expression for
$p_L^2+p_R^2$.
In particular, the lower right
$d{\rm x}d$ block gives the transformed dual torus, which determines
the new momentum modes $G^{-1}m$.
The momentum modes characterize
the volume much more directly than do the winding modes,
since, as can be seen from the formulas for $p_L$ and
$p_R$, turning on a winding mode generates nontrivial
contributions to the momenta and thus to the energy
which depend on B as well as G.  This makes it difficult to
extract G from the spectrum of winding modes.

Under the O(d,d,Z) transformation which exchanges $n_1$
and $m_1$ taking the first momentum mode to the first winding
mode (leaving all other momentum modes fixed),
$G^{-1}$ becomes (generalizing $\rho\leftrightarrow\tau$
from 2 dimensions)
$$1/4G^{\prime -1}=\pmatrix{
{G_{11}-{\left(BG^{-1}B\right)}_{11}}
&1/2{\left(BG^{-1}\right)}_{12}
&\cdots&1/2{\left(BG^{-1}\right)}_{1d}\cr
1/2{\left(BG^{-1}\right)}_{12}&1/4G^{-1}_{22}
&\cdots&1/4G^{-1}_{2d}\cr
\vdots&\vdots&\ddots&\vdots\cr
1/2\left(BG^{-1}\right)_{1d}&1/4G^{-1}_{2d}
&\cdots&1/4G^{-1}_{dd}\cr}$$

Note that the squared volume det $G^{\prime -1}$ is independent
of B: without the $G_{11}$ term,
the first column is a linear combination of the other columns
so that the contributions of $B_{ij}$ to the
determinant cancel.  That is,
$$(BG^{-1}B)_{11}=B_{1k}G^{-1}_{kl}B_{l1}=(1/2B_{1k}G^{-1}_{kl})(2B_{l1})$$
and $$1/2(BG^{-1})_{1j}=1/2B_{1l}G^{-1}_{lj}
=(1/4G^{-1}_{jl})(2B_{l1})$$
so that $${\rm column}1 = \left({\rm column} l\right)\left(2B_{l1}\right)$$
 We are left with
$${\rm det} G^{\prime-1}=4G_{11}V^*_{2d}$$
where $V^*_{2d}$ is the squared dual volume restriced to the 2,3,...,d
directions (i.e. the determinant of $G^{-1}$ restricted to
these components).  Now $G_{11}=e_{1}^2$ and
$e_{1}^2=1/cos^2{\theta_1}e^{*2}_1$ where $\theta_1$is the
angle between $e_1$ and its dual $e^*_1$. So
$${\rm det}G^{\prime-1}={\rm det}G^{-1}\left(4e_1^4 \right)
\left({{V^*_{2d}e^{*2}_1{\rm cos}^2\theta_1}
\over{V^*_{1d}}}\right)$$
But the last factor
is 1, since the volume of the dual lattice is just the volume
of the sublattice spanned by $e^*_2,...,e^*_d$ times the
component of the remaining dual basis vector $e^*_1$
orthogonal to this sublattice.  Thus exchanging $n_1$ and $m_1$
decreases ${\rm det}G^{-1}$ if $e_1^2 < 1/4$.  By repeating this
procedure, reparameterizing if necessary to render $e_1$
the smallest winding, we can continue increasing
${\rm det} G=1/{\rm detG}^{-1}$
as long as the smallest squared winding is less than 1/2.
Then either det G increases ad infinitim or eventually
all windings exceed 1/$\sqrt2$.

In the latter case, it follows
from a theorem in the geometry of numbers
(a generalization of Blichfeldt's theorem
\ref\clos{P.M. Gruber and C.G. Lekkerkerker, {\it Geometry of Numbers}
, 2nd ed., Elsevier Science Publishers, B.V. (1987) 123})
that
$$\sqrt{{\rm det}G} > V_d$$
where $$V_d={{\pi^{d\over 2}{\left(1/2\sqrt 2\right)^d}}
\over{\Gamma(1+d/2)}}$$
is the volume of a sphere of diameter $1/\sqrt2$ in $R^d$.
According to this theorem, any
region in $R^d$ of volume V greater than $\sqrt{{\rm det} G}$ must contain
two points whose difference is in the lattice determined by G.  Consider
a (d-1)-sphere whose diameter is 1/$\sqrt2$.  After the
above procedure we must
have $\sqrt{{\rm det}G}$ greater than the volume of the sphere
since otherwise the sphere would
contain a winding shorter than 1/$\sqrt2$.  Thus for any d,
every theory is equivalent to one compactified
on a volume exceeding $V_d$.

\newsec{Small-large duality for general toroidal compactifications}

With $A\ne 0$, the zero-mode mass spectrum is given by
$$(1/2)(p_L^2+p_R^2)=
\left(\matrix{n^T&m^T&P^T\cr}\right)
M(G,B,A)\left(\matrix{n\cr
                      m\cr
                      P\cr}\right)$$
where M can be read off the expressions for $p_L$
and $p_R$.

Since, as discussed above, the momentum modes characterize
$G^{-1}$, we would like to transform the momentum modes
orthogonally (i.e. via O(16+d,d,Z))
to other modes on the lattice with small
values of $p_L^2+p_R^2$.  If we transform only one of the
momentum modes, that corresponding to the smallest winding $e_1$,
leaving all other momentum modes fixed, the new dual volume
will be the old dual subvolume for the 2 . . . d directions
multiplied by the component of the transformed
first momentum mode orthogonal to this subvolume. If this new
orthogonal component is smaller than the original orthogonal
component, the transformation will reduce the dual volume, as
occurred for the $n_1\leftrightarrow m_1$ transformation with
$A=0$.  Knowing the transformed mode
$(n,m,P)$ (with $2n\cdot m+P^2=0$ so that the new mode is still
0-norm)
suffices to compute the new inverse metric $G^{\prime -1}$, as
long as such an O(16+d,d,Z) transformation exists.  We choose
$$n=\left(\matrix{n_1\cr
                  0\cr
                \vdots\cr
                  0\cr}\right)$$
so that our new mode will still be orthogonal to the other momentum
modes.

As pointed out in \ref\clos{J-P Serre
{\it A Course in Arithmetic}, Springer-Verlag, New York (1973)},
it is always possible to find a dual $x^*$ to x=(n,m,P) such that
the Lorentzian inner products $x^{*.} x=1$ and $x^{*2}=0$ hold and
that $x^*$ is still orthogonal to the winding modes corresponding
to the other momentum modes (2...d).  (Here we are taking x to be
indivisible since we are interested in small values of
$p_L^2+p_R^2$; we can always divide out any common factor
if necessary.)
The 2...d momentum modes still have the corresponding
winding modes as their duals.
The space orthogonal to the new momentum and
winding modes is a 16-dimensional even
self-dual positive definite lattice, either $\Gamma_8 + \Gamma_8$,
the root lattice for $E_8\times E_8$, or $\Gamma_{16}$, the root lattice
for Spin(32)/$Z_2$.

If upon transforming the first momentum mode, the
orthogonal space switches from one root lattice to the other,
our transformation is not quite Lorentzian and therefore
cannot be absorbed into a background field transformation.
It is, however, an
O(16+d,d) transformation from a theory with the same G,B, and A and
with our original momentum
modes but with the orthogonal space the switched root lattice.
This Lorentz transformation $\it{can}$ be absorbed into the background
fields, giving $G^{\prime -1}$ as described above, since this is
determined by the momentum-mode part of the transformation.  This
is simply a restatement of the isomorphism between $\Gamma_8+\Gamma_8$
with one compactified dimension and $\Gamma_{16}$ with one compactified
dimension noted in [3] and [7].  Thus we can always transform the
first momentum mode to any other 0-norm vector on the lattice in
a way that can be obtained by an
O(16+d,d) transformation on the background fields.

We find $$G^{\prime -1}=\left(\matrix{
(1/2)(P+A^Tn)^2+n_1^2G_{11}+(1/4)z_iG^{-1}_{ij}z_j&
(1/4)G^{-1}_{2k}z_k&\cdots&(1/4)G^{-1}_{dk}z_k\cr
(1/4)G^{-1}_{2k}z_k&(1/4)G^{-1}_{22}&\cdots&(1/4)G^{-1}_{2d}\cr
\vdots&\vdots&\ddots&\ddots\cr
(1/4)G^{-1}_{dk}z_k&(1/4)G^{-1}_{d2}&\cdots&(1/4)G^{-1}_{dd}\cr}
\right)$$
where $$z=m-(2B+(1/2)AA^T)n-AP$$
The (1,1) component
of the new dual metric is of course the $(1/2)(p_L^2+p_R^2)$ value for our
new momentum mode; the 2 . . . d components
are those from the original dual metric, since we left the
corresponding momentum modes alone. The others were obtained
by transforming $M(G,B,A)\rightarrow O^TMO$, where O
transforms the first momentum mode to the mode (n,m,P),
leaving the others fixed.

As in the A=0 case, we find that
the first column contains a linear combination of
the others (multiply column j by $z_j$).  When we subtract this
from the first column (since we are taking the determinant), the
first row of the resulting matrix contains the same linear combination
of the other rows as we had above for columns.  Subtracting
this out, the resulting
dual volume is $$\rm{det}(1/4)G^{\prime -1}=
[(1/2)(P_I+A^1_In_1)^2+n_1^2G_{11}](1/4)V^*_{2...d}(1/4)^{d-1}
+z_1^2\rm{det}(1/4)G^{-1}$$

So

$${\rm{det}(1/4)G^{\prime -1}\over\rm{det}(1/4)G^{-1}} =
e_1^2[(1/2)(P+A_I^1n_1)^2+n_1^2e_1^2]
+(m_1-(1/2)A_1^JA_J^1n_1-A^J_1P_J)^2$$

We need to find $n_1$, $m_1$, and P to render this ratio smaller
than 1 as long as $e_1$ remains smaller than some lower bound.  We
can enforce the 0-(Lorentzian) norm condition by taking
$n_1=2y^2$, $P_I=2S_Iy$ and $m_1=-S^2$
where y is an integer and S is a vector in the root lattice.
This ensures that $2n_1m_1+P^2=0$, with $m_1$ integral.
With this choice we have
$$(P+A^1n_1)^2=(2Sy+A^1(2y^2))^2=4y^2(S+A^1y)^2$$
$$=4y^2(D_I+A^{\prime 1}_Iy)\Gamma^{IJ}(D_J+A^{\prime 1}_Jy)$$
and
$$(m_1-(1/2)A_1^JA_J^1n_1-A_1P)^2=\bigl((S+A^1y)^2\bigr)^2$$
$$=\bigl((D_I+A^{\prime 1}_Iy)\Gamma^{IJ}(D_J+A^{\prime 1}_Jy)
\bigr)^2$$
where the $D_I\epsilon Z$ give S in
the integer basis: $S=D_IE_I$ where
the $E_I$ are the basis vectors for the root lattice.  Here $\Gamma$
is the metric for the root lattice: $E_I\cdot E_J=\Gamma_{IJ}$.
Similarly $A^1=A^{\prime 1}_IE_I$.

By a result of diophantine approximation \ref\clos{JWS Cassels,
{\it An Introduction to Diophantine Approximation}, Cambridge
University Press (1965) 13}, it is possible to choose integers
$D_I$ and y such that

$$\vert D_I+A^{\prime 1}_Iy\vert < (\sqrt{e_1})^{1/16}$$
with
$$y<(1/\sqrt{e_1})$$

Then $$e_1y\vert D_I+A^{\prime 1}_Iy\vert <  (\sqrt{e_1})^{17/16}$$
and $$n_1^2e_1^4 < 4e_1^2$$
So $${\rm{det}G^{\prime -1}\over \rm{det}
G^{-1}} < f(e_1)$$ where $f(e_1)$is a positive polynomial in
$e_1^{1/16}$ since the root lattice metric $\Gamma$ is
positive definite.

Setting this polynomial equal to 1 gives a bound on $e_1$
below which the transformation lowers the dual volume,
and thus raises the volume.  Once again, if the smallest
winding $e_1$ never exceeds this finite lower bound, the
volume diverges by repeated transformations of the corresponding
momentum mode.

\newsec{Conclusion}

We have seen that by repeatedly transforming the momentum mode
on the smallest circle, we can prevent the
volume of the compactified dimensions from getting too small.
In two compactified dimensions without Wilson lines,
this transformation is
sufficient in fact to raise all winding lengths.
For any d, if it turns out that the winding lengths
cannot be raised by O(16+d,d,Z), then we can at least say
that the corresponding volume becomes unbounded.  If this
is the whole story, the fact that $\sqrt{{\rm det} G} > V_d $
will imply that some sort of uncertainty
principle is operating on the winding lengths, requiring at
least one of them to blow up if any approach zero.
Since the theory has infinitely many chances to escape
this volume divergence as we iteratively perform
our transformation on the first momentum mode, and since
in two dimensions without Wilson lines windings are bounded
below, we expect windings to have a lower bound in all
cases.

This equivalence between large and small spaces
is often taken to account for the necessary imprecision
in measurements made by fundamental strings of nonzero (Planckian)
size, since it keeps strings insensitive to arbitrarily small
distances.  This does not completely resolve the problem, however,
since duality does not prevent one from measuring {\it exactly}
the volumes of the large and small spaces.
With a Planck-sized ruler one could certainly distinguish a tiny
space from a huge one; what one could not do is measure precisely
the size of either.  The fact that
they cannot be distinguished results directly from the
presence of the winding modes.

Further work is necessary to determine whether windings can
be raised in all cases and the relation of small-large
duality to the modular invariance of the worldsheet.

\newsec{Acknowledgements}
I would like to thank C. Vafa for many helpful discussions and for
introducing me to the subject of small-large duality. I would like
to thank B. Gross
for a helpful discussion on the automorphisms of the lattice. I am
grateful to the Rowland Fund and NSF grant PHY-87-14654 for support.

\listrefs
\bye